\pdfoutput=1

\documentclass[11pt]{article}

\usepackage[final]{acl}

\usepackage{times}
\usepackage{latexsym}
\usepackage[utf8]{inputenc} 
\DeclareUnicodeCharacter{FF0C}{,} 
\usepackage[T1]{fontenc}

\usepackage[utf8]{inputenc}

\usepackage{microtype}
\usepackage{times}  
\usepackage{helvet}  
\usepackage{courier}  
\usepackage{graphicx} 
\usepackage{tabularx}
\usepackage{booktabs}
\usepackage{amsmath}
\usepackage{mathtools}
\usepackage{amssymb} 
\usepackage{natbib}  
\usepackage{caption} 
\usepackage{multirow}
\usepackage{algorithm}
\usepackage{algorithmic}
\usepackage{bibentry}
\usepackage{inconsolata}

\usepackage{graphicx}
\usepackage[utf8]{inputenc}
\usepackage{subcaption}

\title{Optimizing Question Semantic Space for Dynamic Retrieval-Augmented Multi-hop Question Answering}

\author{Linhao Ye, Lang Yu, Zhikai Lei, Qin Chen\textsuperscript{${*}$}, Jie Zhou\textsuperscript{${*}$}, \and Liang He \\
        School of Computer Science and Technology, East China Normal University\\
        \{lhye,lyu,kausal\}@stu.ecnu.edu.cn \{qchen, jzhou, lhe\}@cs.ecnu.edu.cn\\
        \thanks{$^{*}$Corresponding Author.}}

\makeatletter
\def\thanks#1{\protected@xdef\@thanks{\@thanks
\protect\footnotetext{#1}}}
\makeatother

\begin{document}
\maketitle
\begin{abstract}

Retrieval-augmented generation (RAG) is usually integrated into large language models (LLMs) to mitigate hallucinations and knowledge obsolescence. Whereas, conventional one-step retrieve-and-read methods are insufficient for multi-hop question answering, facing challenges of retrieval semantic mismatching and the high cost in handling interdependent subquestions. In this paper, we propose Optimizing Question Semantic Space for Dynamic Retrieval-Augmented Multi-hop Question Answering (\texttt{Q-DREAM}). \texttt{Q-DREAM} consists of three key modules: (1) the Question Decomposition Module (\textit{QDM}), which decomposes multi-hop questions into fine-grained subquestions; (2) the Subquestion Dependency Optimizer Module (\textit{SDOM}), which models the interdependent relations of subquestions for better understanding; and (3) the Dynamic Passage Retrieval Module (\textit{DPRM}), which aligns subquestions with relevant passages by optimizing the semantic embeddings.
Experimental results across various benchmarks demonstrate that \texttt{Q-DREAM} significantly outperforms existing RAG methods, achieving state-of-the-art performance in both in-domain and out-of-domain settings. Notably, \texttt{Q-DREAM} also improves retrieval efficiency while maintaining high accuracy compared with recent baselines.
\end{abstract} 


\begin{figure}[ht]
    \centering
    \begin{subfigure}[b]{0.5\textwidth}
        \centering
        \includegraphics[width=\textwidth]{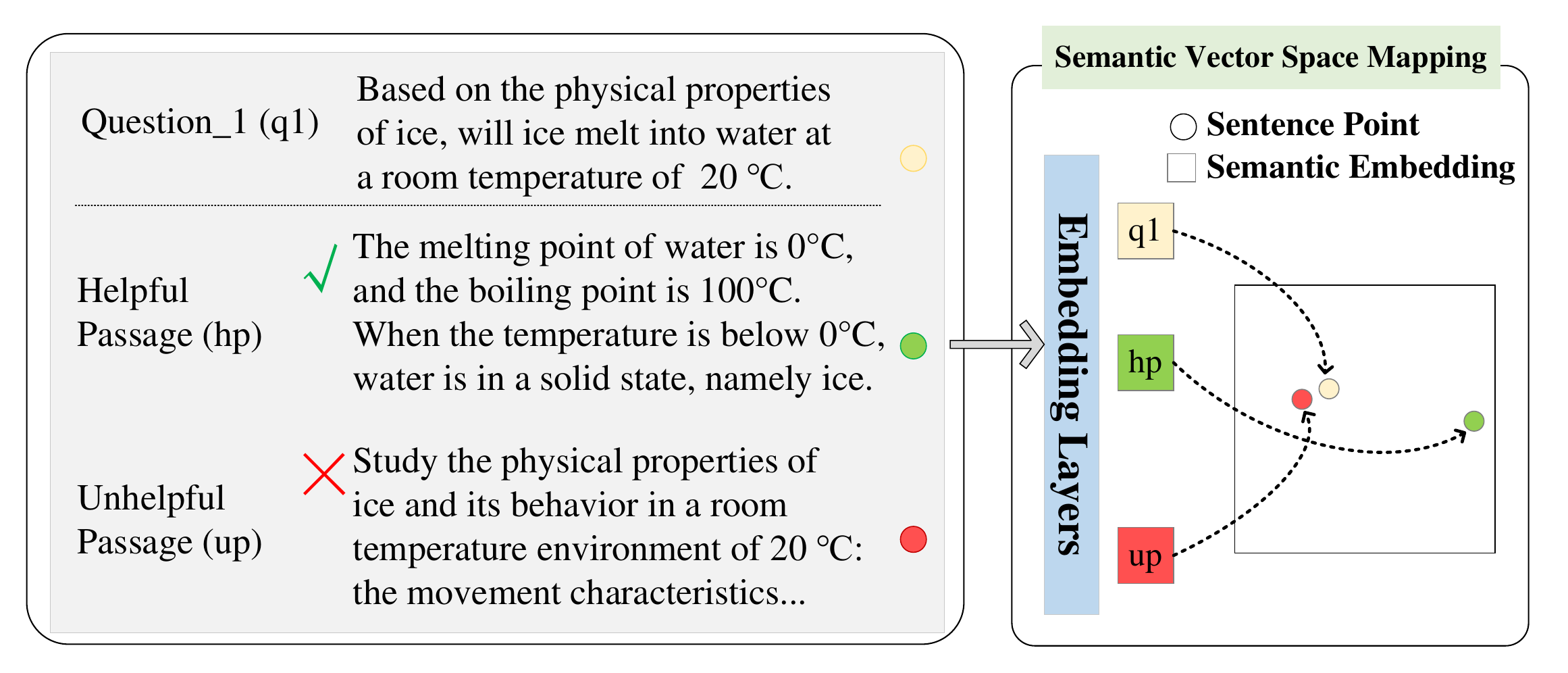} 
        \caption{Semantic Mismatching}
        \label{fig:sub1}
    \end{subfigure}
    \vskip 0.1em 

    \begin{subfigure}[b]{0.5\textwidth}
        \centering
        \includegraphics[width=\textwidth]{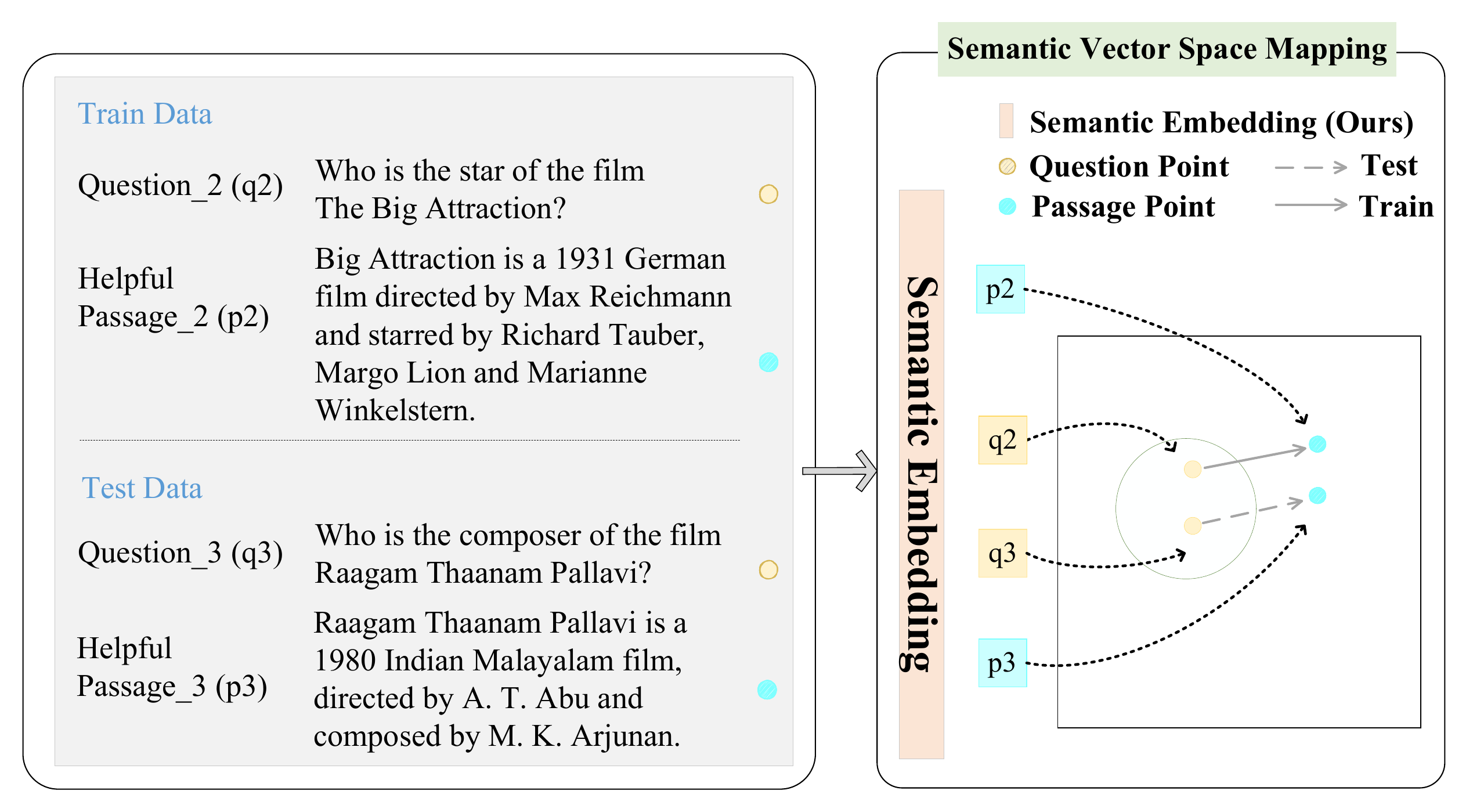} 
        \caption{Semantic Space Optimization}
        \label{fig:sub2}
    \end{subfigure}

    \caption{(a) Semantic Mismatching: A semantic proximity gap leads to retrieval failures between question and helpful passage. (b) Semantic Space Optimization: Semantic embedding optimization aligns question and helpful passage by learning the latent semantic matching pattern.}
    \label{fig:main}
\end{figure}

\section{Introduction}

Recently, the advent of Large Language Models (LLMs), such as GPT \citep{chatgpt}, LLaMA \citep{llama2}, and Mistral \citep{mistral-7b}, has significantly expanded the boundaries of machine language understanding and generation, enhancing the performance of a wide range of NLP tasks \citep{bang2023multitask, ouyang2022training}. However, they also tend to exhibit an inclination to generate hallucinations \cite{bang2023multitask, guerreiro2023hallucinations, chen2024diahalu}. In addition, LLMs inherently suffer from knowledge obsolescence, as they are trained on static datasets collected at a fixed point in time \cite{dhingra2022time, huang2020challenges}. Consequently, the responses do not incorporate real-time updates or newly emerging information, which can be critical for many real-world applications \citep{zhang2024mme, nguyen2024sfr, chen2025enhancing}. 

One solution is to periodically retrain these models on updated corpora, while this approach is both computationally expensive and time-consuming \citep{jiang2024megascale, xia2024understanding}. A more efficient alternative is retrieval-augmented generation (RAG), which integrates LLMs with information retrieved from external updated knowledge bases \citep{ram2023context, ye2024boosting, guu2020retrieval, shi2023replug}. Whereas, the RAG methods struggle with multi-hop QA tasks, as they usually fail to handle combinatorial questions requiring information from multiple passages. Recent prompting-based approaches \citep{selfask, decomposed, sure} like IRCoT \citep{ircot} attempt to address the multi-hop QA tasks by interleaving retrieval with chain-of-thought (CoT) reasoning. While IRCoT leverages retrieved results to refine reasoning, it heavily relies on the CoT capability of the model and demands multiple interactions between retriever and generator, incurring high computational costs.

  Another challenge is that the existing RAG methods suffer from the semantic mismatching problem, which introduces the semantic similar but unhelpful passages for generation and negatively impacts the performance of multi-hop QA. In other words, even semantically similar passages may lack relevance to the question, leading models to prioritize the high similar but unhelpful content. 
  As shown in Figure \ref{fig:sub1}, for the question "will ice melt into water at a root temperature at 20°C?", the method incorrectly retrieves a passage describing the movement characteristics of ice (high similar but unhelpful), while ignoring the low similar passage containing the relevant fact "The melting point of water is 0°C". This mismatching problem stems from the semantic proximity gap between the question and the truly helpful passages.



To address the above problems, we propose Optimizing \textbf{Q}uestion Semantic Space for \textbf{D}ynamic \textbf{Re}trieval-\textbf{A}ugmented \textbf{M}ulti-hop Question Answering (\texttt{Q-DREAM}), which enhances retrieval efficiency and resolves semantic mismatching problem with a three-module pipeline. 
The modules consist of the Question Decomposition Module (\textit{QDM}), Subquestion Dependency Optimizer Module (\textit{SDOM}), and Dynamic Passage Retrieval Module (\textit{DPRM}). The \textit{QDM} module first decomposes complex questions into fine-grained subquestions. Each subquestion is then processed separately. The independent subquestions are directly passed to \textit{DPRM} for retrieval, while dependent subquestions are refined by \textit{SDOM} before retrieval. The \textit{DPRM} module integrates semantic alignment mechanism, which clusters subquestions and maps each cluster to a dedicated retrieval space for dynamic retrieval. 
As illustrated in Figure \ref{fig:sub2}, our method bridges the semantic gap between the question and the helpful passage. During training, the semantic embeddings of question q2 and helpful passage p2 are optimized to be closer in the dedicated retrieval space, which captures the latent matching pattern by associating "[Role] of [Film]" with the helpful content ("Film is..., Role by...").
At test time, the similar question q3 is mapped to the same cluster as q2. As both questions share the same semantic pattern "[Role] of [Film], q3 can align with the helpful passage p3 (Raagam Thaanam Pallavi is..., composed by...") by leveraging the learned matching patterns during training. 

We perform extensive experiments on various datasets, and the results demonstrate that \texttt{Q-DREAM} significantly outperforms existing approaches in handling multi-hop questions, achieving superior performance across in-domain and out-of-domain settings.
The main contributions can be summarized as follows:
\begin{itemize}
    \item We propose a novel retrieval-augmented framework, namely \texttt{Q-DREAM}, which is model-agnostic and can be easily adapted to various LLMs to enhance the retrieval efficiency and effectiveness for retrieval-augmented multi-hop QA.
    \item Three modules as \textit{QDM}, \textit{SDOM} and \textit{DPRM} are integrated into the framework, which work collaboratively to address the reconstruction of interdependent subquestions and resolve the semantic mismatching issues.
    \item  We conduct elaborate analyses of the experimental results on three benchmark datasets, demonstrating the effectiveness of \texttt{Q-DREAM} under both in-domain and out-of-domain settings, and exhibiting scalability across various LLM backbones. 
\end{itemize}

\section{Related Work}
\subsection{Task Decomposition}

Task decomposition is a crucial approach for addressing complex tasks, particularly in multi-turn and multi-hop question answering. Prior studies have explored various methods to break down complex questions into a series of simpler subquestions. Several works \citep{iyyer2017search,talmor2018web,rao2019answer,wolfson2020break,khot2022decomposed} propose models that decompose complex questions, yet these approaches do not leverage pre-trained language models (LMs). More recent methods, such as \citep{wang2022shepherd}, utilize pre-trained models to generate contextual information for multiple-choice tasks. In addition, SEQZERO \citep{yang2022seqzero} introduces a few-shot semantic parsing technique that decomposes questions into structured subquestions aligned with a formal representation, enabling efficient reasoning through concise prompts. RA-ISF \citep{raisf} further refines decomposition strategies by iteratively answering subquestions to minimize the impact of irrelevant text. However, despite these advancements, existing question decomposition techniques often fail to adequately handle interdependent subquestions in retrieval-augmented settings. Since the retrieval of each subquestion is performed independently, it may lead to incomplete or suboptimal retrieval results, reducing the accuracy of multi-hop question answering. 



\subsection{Retrieval-Augmented Language Models}

Retrieval-augmented language models (LMs) enhance the reasoning and factual accuracy of LMs by incorporating externally retrieved information, thereby mitigating hallucination and factual inconsistency issues in open-domain question answering (ODQA) \citep{guu2020retrieval,lewis2020retrieval,lazaridou2022internet}. In previous studies, such as REALM \cite{guu2020retrieval} jointly optimizes the retriever and language model to enhance retrieval-aware generation, RETRO \cite{borgeaud2022improving} introduces training language models on top of a frozen retriever, Atlas \cite{izacard2022few} advances further by exploring dedicated loss functions for end-to-end training of both the retriever and the LM, demonstrating superior performance in few-shot learning tasks, RePlug \cite{shi2023replug} maintains a frozen black-box LM during the fine-tuning of retrieval modules. 

However, these retrieval-augmented approaches struggle with multi-hop reasoning in QA. Recently, researchers have explored prompt-based methods to improve multi-hop QA. SelfAsk \citep{selfask} enhances retrieval by integrating structured prompting and search engines. DecomP \citep{decomposed} decomposes tasks into modular sub-prompts tailored to specific reasoning steps. SURE \citep{sure} employs prompts to guide the LLMs in generating summaries for each answer candidate. All of these approaches do not utilize the Chain-of-Thought (CoT) reasoning of LLMs. ReAct \citep{react} combines reasoning and action, prompting LLMs to generate task-related reasoning traces and actions in an interactive manner. IRCoT \citep{ircot} integrates retrieval with CoT reasoning, using the reasoning to guide the retrieval and then leveraging the retrieval results to refine the reasoning process. However, such interactive approaches rely heavily on model performance and introduce a high computational cost.

\section{Method}
\begin{figure*}[h]
    \centering
    \includegraphics[width=\linewidth]{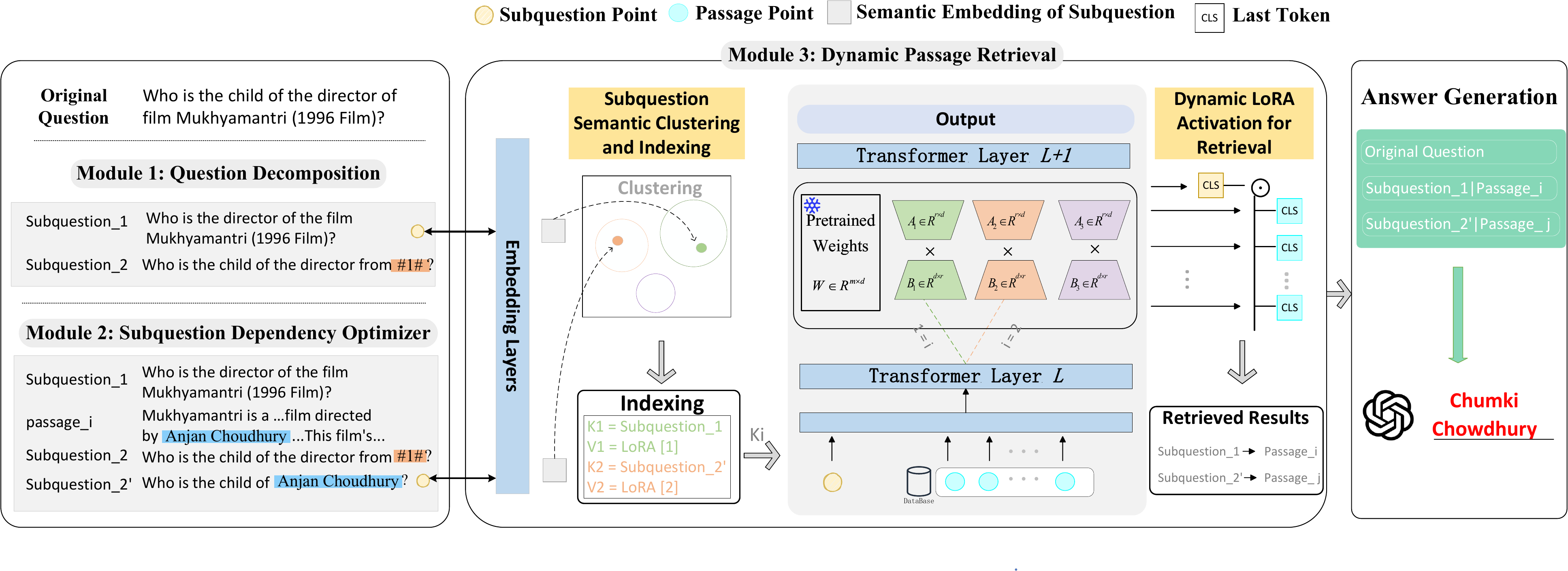}
    \caption{Overall framework of \texttt{Q-DREAM}. }
    \label{Method}
\end{figure*}

\subsection{Overview}

The overall architecture of our approach is shown in Figure \ref{Method}, which consists of three modules: Question Decomposition Module (\textit{QDM}), Subquestion Dependency Optimizer Module (\textit{SDOM}) and Dynamic Passage Retrieval Module (\textit{DPRM}). \textit{QDM} first decomposes complex questions into fine-grained subquestions. These subquestions then undergo individual processing, where independent ones proceed directly to \textit{DPRM} for retrieval while dependent ones are refined by \textit{SDOM} before retrieval. \textit{DPRM} incorporates a semantic alignment mechanism that clusters subquestions and maps each cluster to a dedicated retrieval space for dynamic retrieval.

Specifically, an origin question as $Q_{ori}$ is input into the \texttt{Q-DREAM} framework to obtain the answer. The overall process is as follows: Firstly, we use the \textit{QDM} to decompose $Q_{ori}$ into subquestions $Q_{sub} = \{q_1, \ldots, q_n \}$. If a subquestion $q_i$ does not depend on the answer to a previous subquestion, it is directly sent to the \textit{DPRM} for passage retrieval. Otherwise, $q_i$ is first optimized by the \textit{SDOM} to generate a new subquestion ${{q_i}\prime}$, which is then sent to the \textit{DPRM}.


In the \textit{DPRM}, each subquestion ${q_i}^*$ (where ${q_i}^*$ is either $q_i$ or ${{q_i}\prime}$) is first assigned to a semantic cluster based on its embedding. A corresponding LoRA block is then indexed and used for retrieval according to the cluster labels. Next, ${q_i}^*$ and the candidate retrieved passages $P = \{p_1, \ldots, p_m \}$ are encoded with the corresponding LoRA block. The embeddings $E_{{q_i}^*}$ and $E_P = \{E_{p_1}, \ldots, E_{p_m} \}$ are extracted from the last hidden state of the last layer of the model. We calculate the similarity between $E_{{q_i}^*}$ and each $E_{P_v}$, and the passage with the highest similarity score is selected as the retrieved result for ${q_i}^*$. Finally, the original question, the subquestions, and their corresponding retrieved passages are integrated into the answer generation model to generate the final answer.




\subsection{ Q-DREAM based Training}
In this section, we will introduce the training process for the three modules of \texttt{Q-DREAM}.



Question Decomposition Module (\textit{QDM}) decomposes $Q_{ori}$ into subquestions via a model $M_{QDM}$. For training the model, we optimize:
\begin{equation}
    \max_{M_{QDM}} \log P(Q_{sub} \mid Q_{ori}; M_{QDM})
\end{equation}
where $Q_{sub} = \{q_1, \ldots, q_n \}$, and $q_i$ is a subquestion.

Subquestion Dependency Optimization Module \textit{(SDOM)} aims at refining the subquestions that have dependencies on others. 
For a subquestion $q_i$ that requires the answer to $q_j$, \textit{SDOM} optimizes $q_i$ by utilizing the retrieved passage $p_{q_j}$ of $q_j$ as:
\begin{equation}
\max_{M_{SDOM}} \log P({{q_i}\prime} \mid q_j,p_{q_j},q_i; M_{SDOM})
\end{equation}

Dynamic Passage Retrieval Module (\textit{DPRM}) integrates semantic alignment mechanism to address the issue of semantic mismatching. We cluster subquestions into multiple categories and train cluster-specific embeddings to align questions and helpful passages in the semantic space. Given subquestions processed through \textit{QDM} and \textit{SDQM}, we partition them into k clusters $\{C_1, \ldots, C_k \}$ using the k-means algorithm on their embeddings. For each cluster $C_i$, we fine-tune a LoRA block $M_{DPRM}^{C_i}$ to maximize the similarity between the subquestion $q \in {C_i}$ and the helpful passage p:
\begin{equation}
\max_{M_{DPRM}^{C_i}} \mathbb{E}_{(q, p) \in C_i} \log \left( \frac{\mathbf{E}_q \cdot \mathbf{E}_p}{\|\mathbf{E}_q\| \|\mathbf{E}_p\|} \right),
\end{equation}
where $E_q$ and $E_p$ are the embeddings from the final hidden state of the last layer of $M_{DPRM}^{C_i}$.

\subsection{Q-DREAM based Inference}
In this section, we offer a comprehensive overview of how \texttt{Q-DREAM} framework processes and generates an answer for the original question $Q_{ori}$. Algorithm \ref{alg:Q-DREAM_inference} presents the details of \texttt{Q-DREAM}. During the inference, we use all pre-trained models of the three modules.

\textbf{Question Decomposition. } In this process, we employ the $M_{QDM}$ to decompose the original question $Q_{ori}$ into multiple subquestions $Q_{sub}$, which is formulated as follows:
\begin{equation}
     \underset{Q_{sub}}{\arg \max} \ P(Q_{sub} \mid Q_{ori}; M_{QDM})
\end{equation}

Next, we process the subquestions in $Q_{sub}$ sequentially. If the subquestion $q_i$ does not depend on the answer to previous subquestion, then directly input to the \textit{DPRM} and retrieve the passage $p_{q_i}$. Otherwise, $q_i$ is sent to the next step. 

We determine whether a subquestion exhibits dependencies by checking if there is a "\#number\#" marker by itself.
As shown in Figure \ref{Method}, the decomposed Subquestion\_2 "Who is the child of the director from \#1\#?
" exhibits dependencies, with the marker "\#1\#" indicating that it depends on the answer to Subquestion\_1.

\begin{algorithm}[h]
\caption{The inference process of Q-DREAM }
\label{alg:Q-DREAM_inference}
\begin{algorithmic}[1]
\REQUIRE $M, M_{QDM}, M_{SDOM}, M_{DPRM},$ \\
\vspace{1mm}
\hspace*{8.0mm} $Q_{ori}, {Q_{sub}\prime} \leftarrow \emptyset, {P_{Q_{sub}\prime}} \leftarrow \emptyset$ \\
\vspace{1mm}
\ENSURE $A_{ori}$
\vspace{1mm}
\hrule
\vspace{1mm}
\STATE $Q_{sub} = \{q_1, \ldots, q_n\} \leftarrow M_{QDM}(Q_{ori})$
\FOR{$i = 1$ to $n$}
    \IF{$q_i$ \text{depend on the answer to }$q_j$}
        \STATE ${{q_i}\prime} = M_{SDOM}(q_j, p_{q_j}, q_i)$
        \STATE ${{q_i}^*} = {{q_i}\prime}$
    \ELSE
        \STATE ${{q_i}^*} = {q_i}$
    \ENDIF
    \STATE${Q_{sub}\prime} \leftarrow 
    {Q_{sub}\prime} \cup \{{{q_i}^*}\}$
    \STATE$c_{{q_i}^*} = cluster({{q_i}^*})$, \text{Max\_Score = -1}
    \STATE$E_{{q_i}^*} = h_{Last}^{(M)} \leftarrow {M_{DPRM}^{c_{{q_i}^*}}}({{q_i}^*})$
    \FOR{$p_v$ to $P$}
        \STATE $E_{p_v} = h_{Last}^{(M)} \leftarrow {M_{DPRM}^{c_{{q_i}^*}}}({p_v})$
        \STATE $\text{score} = cos(E_{{q_i}^*}, E_{p_v})$ 
        \IF{\text{score $>$ Max\_Score}}
            \STATE \text{Max\_Score = score}
            \STATE $p_{{q_i}^*} = p_v$
        \ENDIF
    \ENDFOR
    \STATE${P_{Q_{sub}\prime}} \leftarrow 
    {P_{Q_{sub}\prime}} \cup \{p_{{q_i}^*}\}$
\ENDFOR

\STATE $A_{ori} = M(Q_{ori}, {Q_{sub}\prime}, {P_{Q_{sub}\prime}})$
\end{algorithmic}
\end{algorithm}

\textbf{Subquestion Dependency Optimizer. }
If subquestion $q_i$ depends on the answer to subquestion \#j\#, it is incomplete and cannot effectively retrieve relevant content. Therefore, we need to utilize subquestion $q_j$ and its retrieved passage $p_{q_j}$ that contains useful information for answering $q_j$, to optimize $q_i$, which is formulated as:
\begin{equation}
    \underset{{{q_i}\prime}}{\arg \max} \ P({{q_i}\prime} \mid q_j, p_{q_j}, q_i; M_{SDOM}) 
\end{equation}
where ${{q_i}\prime}$ is the reconstructed subquestion after optimization.

After optimization, reconstructed subquestion\_2 "Who is the child of Anjan Choudhury?" in Figure \ref{Method} becomes more complete by explicitly identifying 'Anjan Choudhury' as the director during the retrieval process. Moreover, this explicit reconstruction process acts as the Chain-of-Thought, which helps the answer generation model to improve the reasoning performance for multi-hop QA.

\textbf{Dynamic Passage Retrieval. }
In the Dynamic Passage Retrieval Module, each subquestion ${q_i}^*$ (where ${q_i}^*$ is either $q_i$ or ${{q_i}\prime}$) is clustered based on its semantic embedding and the corresponding LoRA block is indexed. Subsequently, ${q_i}^*$ and candidate passages $P = \{p_1, \ldots, p_m \}$ are input into $M^{c_{{q_i}^*}}_{DPRM}$ respectively. We obtain $E_{{q_i}^*}$ and $E_P = \{E_{p_1}, \ldots, E_{p_m} \}$ form the last hidden state of the last layer of $M^{c_{{q_i}^*}}_{DPRM}$ and select the passage with the highest score according to the following formula:

\begin{equation}
p_{{q_i}^*} = \underset{p_v \in P}{\arg \max} \
 f({E_{{q_i}^*}}, E_{p_v})
\end{equation}
where $f(\cdot)$ is a score function such as cosine similarity, and $p_{{q_i}^*}$ is the retrieved passage of ${q_i}^*$.


Finally, we replace the subquestions in $Q_{sub}$ that exhibit dependencies with those reconstructed from \textit{SDOM}, forming ${{Q_{sub}}\prime}$, and obtain the retrieved passages $P_{{{Q_{sub}}\prime}}$ with \textit{DPRM}. Then the original question $Q_{ori}$, the subquestions ${{Q_{sub}}\prime}$, and the retrieved passages $P_{{{Q_{sub}}\prime}}$ are input into the answer generation model $M$ to generate the answer, which is formulated as follows:
\begin{equation}
    \underset{{A_{ori}}}{\arg \max} \ P(A_{ori} \mid Q_{ori}, {{Q_{sub}}\prime}, P_{{{Q_{sub}}\prime}}; M)
\end{equation}
where $A_{ori}$ is the answer to the original question.

\begin{table*}[h]
    \centering
    \caption{EM / F1 scores for different methods on three datasets. \textbf{Bold} number indicates the best performance among all methods. * indicates the results from the original paper.}
    \label{tab:main_results}
    \setlength{\extrarowheight}{5pt}
    \setlength{\tabcolsep}{15pt}
    \begin{tabularx}{1\textwidth}{l|X|XX|c}
        \toprule
        \multirow{2}{*}{Methods/Datasets} & \multicolumn{1}{c|}{In-domain} & \multicolumn{2}{c|}{Out-of-domain} & \multirow{2}{*}{\centering Average}\\ 
        \cmidrule{2-4} 
        & \multicolumn{1}{c|}{2WikiMQA} & \multicolumn{1}{c}{HotpotQA} & \multicolumn{1}{c|}{IIRC} &  \\
        \midrule
        \multicolumn{5}{@{}l@{}}{\textbf{Llama}} \\ 
        \midrule
        \mbox{InstructRAG (Llama3-8B)} &  \multicolumn{1}{c|}{30.4/38.9} & \multicolumn{1}{c}{22.6/32.1} & \multicolumn{1}{c|}{14.2/18.1} & \multicolumn{1}{c}{22.4/29.7}\\
        ChatQA2 (Llama3-8B) & \multicolumn{1}{c|}{29.0/35.2} & \multicolumn{1}{c}{32.6/42.5} & \multicolumn{1}{c|}{21.4/25.7} & \multicolumn{1}{c}{27.7/34.5}\\
        \texttt{Q-DREAM} (Llama2-7B) & \multicolumn{1}{c|}{\textbf{32.4/39.7}} & \multicolumn{1}{c}{\textbf{36.0/46.1}} & \multicolumn{1}{c|}{\textbf{23.8/27.2}} & \multicolumn{1}{c}{\textbf{30.7/37.7}}\\
        \midrule
        \multicolumn{5}{@{}l@{}}{\textbf{ChatGPT}} \\ 
        \midrule
        ChatGPT &  \multicolumn{1}{c|}{23.6/29.2} & \multicolumn{1}{c}{24.2/32.5} & \multicolumn{1}{c|}{11.4/13.8} & \multicolumn{1}{c}{19.7/25.2}\\
        SURE & \multicolumn{1}{c|}{32.8*/38.1*} & \multicolumn{1}{c}{33.2*/43.4*} & \multicolumn{1}{c|}{20.6/25.5} & \multicolumn{1}{c}{28.9/35.7}\\
        IRCoT & \multicolumn{1}{c|}{41.9/55.2} & \multicolumn{1}{c}{25.5/38.1} & \multicolumn{1}{c|}{21.0/31.0} & \multicolumn{1}{c}{29.5/41.4}\\
        \texttt{Q-DREAM} & \multicolumn{1}{c|}{\textbf{48.6/62.1}} & \multicolumn{1}{c}{\textbf{48.4/60.9}} & \multicolumn{1}{c|}{\textbf{28.2/31.9}} & \multicolumn{1}{c}{\textbf{41.7/51.6}}\\
        \bottomrule
    \end{tabularx}
\end{table*}


\section{Experimental Setup}
\subsection{Datasets}
We use the following three multi-hop QA datasets in the open-domain setting to evaluate \texttt{Q-DREAM}: \textbf{HotpotQA} \cite{hotpotqa}, \textbf{2WikiMultihopQA} (2WikiMQA) \citep{2wiki}, \textbf{IIRC} \citep{iirc}. For the above three datasets, we use the subsampled splits released by \citep{ircot} as our test set. To evaluate the generalization of \texttt{Q-DREAM}, we only utilize 2WikiMQA (in-domain) for training, and HotpotQA and IIRC as the out-of-domain benchmarks for testing.


\subsection{Baselines and Evaluation Metrics}
We compare with the recent advanced baselines:

\textbf{InstructRAG} \citep{wei2025instructrag}: allows LMs to denoise retrieved contents by generating rationales for better verifiability and trustworthiness.

\textbf{ChatQA2} \citep{xu2024chatqa}: bridge the gap between open-source LLMs and leading proprietary models in long context understanding and retrieval-augmented generation capabilities.

\textbf{ChatGPT} \citep{chatgpt}: excels at multi-hop questions by leveraging its ability to understand and analyze questions. In the experiment setting, we evaluate ChatGPT in a one-shot prompting setting to guide its reasoning process.

\textbf{SURE} \citep{sure}: enhances question-answering tasks by summarizing retrieved passages and selecting the most plausible answer from multiple candidates.

\textbf{IRCoT} \citep{ircot}: interleaves retrieval with CoT reasoning, dynamically refining the retrieval process based on intermediate reasoning steps.

We evaluate the QA performance using the standard exact match (EM) and F1 scores, which have been widely used in previous studies \cite{ircot, sure}.

\subsection{Implementation Details}
Our framework consists of one answer generation model and three modules that serve as intermediate components. For the answer generation model, we experiment with open-source Llama2-7B \citep{llama2} and closed-source ChatGPT (GPT-3.5-turbo) \citep{chatgpt} through the API. As for the three modules, we fine-tune Mistral-7B \citep{mistral-7b} for \textit{QDM} and \textit{SDOM}, and E5-Mistral-7B-Instruct (E5-Mistral) \citep{e5-mistral-7b} for \textit{DPRM}, which is pre-trained on a multilingual mixed data set.

We initially select 15,000 samples from the 2WikiMQA and generate training labels for \textit{QDM} and \textit{SDOM} using ChatGPT. After data cleaning and denoising, we ultimately obtain 13,363 samples as our training set. In the \textit{DPRM}, we use the k-means \citep{kmeans} algorithm for clustering. During training, we adopt Adam \citep{adam} with a constant learning rate of 5e-5 and a dropout rate of 10\%. we set the batch size to 16 and train the model with one A100. We use greedy decoding during the inference process across all experiments to ensure deterministic generation. 

\section{Results and Analyses}
\subsection{Main Results}
We report the EM and F1 scores on three multi-hop QA datasets under in-domain and out-of-domain settings, and the results are shown in Table \ref{tab:main_results}. 

\textbf{(1) In-Domain Performance with ChatGPT.} \texttt{Q-DREAM} achieves state-of-the-art performance on 2WikiMQA with ChatGPT as the backbone model. Compared to ChatGPT without retrieval augmentation, our method exhibits absolute improvements of 25.0 and 32.9 regarding the EM and F1 scores respectively. Moreover, \texttt{Q-DREAM} outperforms the strongest retrieval-augmented baseline IRCoT with substantial improvements. These findings verify the effectiveness of our method in multi-hop question answering.

\textbf{(2) Generalization in Out-of-Domain Settings.} To evaluate the generalization of our method, we directly apply our method trained on 2WikiMQA to HotpotQA and IIRC. As shown in the "Out-of-domain" column of Table \ref{tab:main_results}, \texttt{Q-DREAM} surpasses all baselines in out-of-domain datasets, achieving great improvements of 17.5 in terms of F1 compared with the second best model SURE in HotpotQA.
On the whole, our method is superior to all the baselines in the average performance, demonstrating good generalization and robust cross-domain adaptability of our method.

\textbf{(3) Performance with Smaller-Scale Model.}
To show the effectiveness of our method with smaller-scale models, we also experiment \texttt{Q-DREAM} with Llama2-7B as the backbone. The results show that \texttt{Q-DREAM} achieves an average performance of 30.7 and 37.7 in EM and F1 scores respectively across three datasets, surpassing baselines built on larger architectures like InstructRAG and ChatQA2. Moreover, \texttt{Q-DREAM} with Llama2-7B even outperforms SURE that uses larger-scale ChatGPT as the backbone, indicating the superiority of our RAG method especially for smaller LLMs.



\subsection{Ablation Studies}
We set up three variants on 2WikiMQA to investigate the effectiveness of each component of \texttt{Q-DREAM}:
\begin{itemize}
    \item \textit{- DPRM}: This variant refers to the subquestions generated by the \textit{QDM} and \textit{SDOM} being directly input to the E5-Mistral model for passage retrieval, without utilizing the \textit{DPRM}.
    \item \textit{- (SDOM, DPRM)}: After \textit{QDM}, the question is decomposed into subquestions which are subsequently directly input to the E5-Mistral model for passage retrieval.
    \item \textit{- All}: The question is directly input to the E5-Mistral model for passage retrieval. For a fair comparison, the number of passages retrieved in this variant is consistent with our method.
\end{itemize}



The results are shown in Table \ref{tab:Ablation}. We can see that removing \textit{DPRM} substantially degrades performance, attributable to the absence of semantic matching optimization with dynamic dedicated retrieval. Concurrently removing \textit{SDOM} and \textit{DPRM} yields the most severe deterioration, as directly using the decomposed questions without optimization for conventional retrieval is prone to the semantic mismatching problem. Removing all components achieves higher performance than \textit{- (SDOM, DPRM)} alone, underscoring the critical role of  \textit{SDOM}. Without \textit{SDOM} optimization, the dependent subquestions are vague and lack the necessary contextual information for effective retrieval. Overall, each module within the framework plays an indispensable role in multi-hop QA.


\begin{table}[t]
\scriptsize
\centering
\caption{The results of ablation studies.}
\label{tab:Ablation}
\setlength{\tabcolsep}{3.2pt} 
\resizebox{0.3\textwidth}{!}{
\begin{tabular}{lcc}
\toprule
Methods & EM & F1 \\
\midrule
\texttt{Q-DREAM} & \textbf{48.6} & \textbf{62.1}\\
\textit{- DPRM} & 44.4 & 56.5\\
\textit{- (SDOM, DPRM)} & 32.4 & 38.1\\
\textit{- All} & 37.2 & 44.7\\
\bottomrule
\end{tabular}
}
\end{table}

\begin{table*}
    \centering
    
    \caption{Efficiency of inference on 2WikiMQA. Avg(s)/Dataset represents the average time required to process the dataset. Avg(s)/Sample represents the average time required to process a single sample.}
    \label{tab:speed}
    \setlength{\extrarowheight}{5pt}
    \setlength{\tabcolsep}{15pt}
    \begin{tabularx}{1.00\textwidth}{X|X|X|X|X|c}
        \toprule
        \multirow{2}{*}{Methods} & \multicolumn{3}{c|}{Inference Time (s)} & \multirow{2}{*}{\centering Avg(s)/Dataset} & \multirow{2}{*}{\centering Avg(s)/Sample}\\ 
        \cmidrule{2-4} 
        & \multicolumn{1}{c|}{First} & \multicolumn{1}{c|}{Second} & \multicolumn{1}{c|}{Third} &  \\
        \midrule
        IRCoT &  \multicolumn{1}{c|}{8887} & \multicolumn{1}{c|}{16138} & \multicolumn{1}{c|}{12485} & \multicolumn{1}{c|}{12503} & \multicolumn{1}{c}{25}\\
        \texttt{Q-DREAM} & \multicolumn{1}{c|}{2009} & \multicolumn{1}{c|}{2000} & \multicolumn{1}{c|}{2025} & \multicolumn{1}{c|}{2011} & \multicolumn{1}{c}{4}\\
        \bottomrule
    \end{tabularx}
\end{table*}

\begin{table*}[h]
\centering
\caption{Comparison of the retrieval performance. - \textit{DPRM} refers to an ablated variant of \texttt{Q-DREAM} without its dynamic passage retrieval module.}
\label{tab:Retrieval_Performance_Comparison}
\begin{tabular}{lcccccc}
\toprule
Metric / Model
 & BM25 & Contriever & InstructRAG & IRCoT & \texttt{Q-DREAM} (- \textit{DPRM}) & \texttt{Q-DREAM} \\
\midrule
Precision (\%) & 41.0 & 20.5 & 19.7 & 53.8 & 57.4 & \textbf{81.8} \\
Recall (\%)    & 68.0 & 33.0 & 40.7 & 68.9 & 62.0 & \textbf{85.7} \\
F1 (\%)        & 51.2 & 25.3 & 26.5 & 60.4 & 59.6 & \textbf{83.7} \\
\bottomrule
\end{tabular}
\end{table*}

\subsection{Further Analyses}

\paragraph{Efficiency of Inference.}


The baseline methods require multiple model interactions during the reasoning process, resulting in substantial time consumption. In contrast, our approach utilizes pre-trained sub-modules to generate fine-grained subquestions and retrieve relevant passages, which are then directly processed by the final question generation model. Thus, our method is more efficient.

To quantify the efficiency of inference, we evaluate the inference time in the 2WikiMQA test set. We compare \texttt{Q-DREAM} with the strong baseline IRCoT on the same computing platform with uniform hardware configurations and operating systems. To ensure the reliability of the results, each method run three times independently under the same conditions.
As shown in Table \ref{tab:speed}, \texttt{Q-DREAM} processes each sample in 4 seconds on average, which is 6× faster than IRCoT. Additionally, our method demonstrates a smaller variance in inference time across three experiments, highlighting its stronger stability compared to IRCoT. These characteristics are crucial for practical applications.


\paragraph{Performance of Retrieval.}
To further assess the effectiveness of our retrieval method, we compare with the classical and recent advanced information retrieval (IR) models. Three categories of baselines are used for comparison: BM25 \citep{jones2000probabilistic} is a lexical-based retrieval model; Contriever \citep{izacard2021unsupervised} is a dense neural retriever; and the retrieval methods integrated in InstructRAG and IRCoT. 
For the space of limit, we present the results of 2WikiMQA in Table \ref{tab:Retrieval_Performance_Comparison}. 

It is observed that our retrieval method with \textit{DPRM} obtains superior retrieval performance compared with the pure IR models and the IR methods used in recent RAG baselines. By removing \textit{DPRM} and using the initial E5-Mistral model for retrieval, the retrieval performance significantly drops since all questions and passages share the same semantic encoder, which is prone to the semantic mismatching problem. By integrating our retrieval method as \textit{DPRM}, the similar questions are assigned with a dedicated LoRA block for retrieval, which forces the embeddings of the helpful passages to be closer to the question in the question semantic space.



\paragraph{Impact of Retrieved Passages.}

\begin{figure}[t]
    \centering
    \begin{minipage}{0.5\textwidth}
        \includegraphics[width=\linewidth]{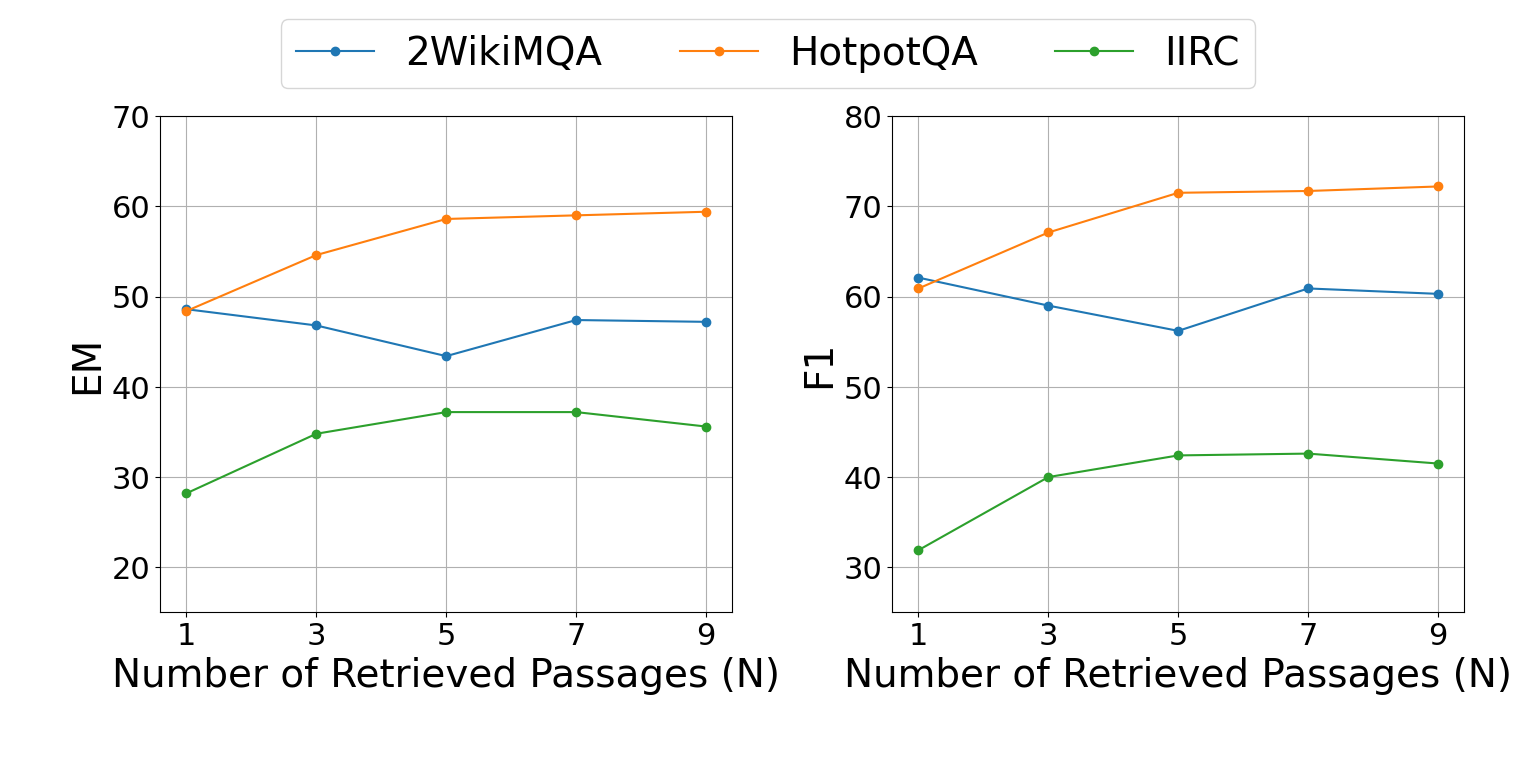}
        \caption{Impact of the number of retrieved passages.}
        \label{figure_retrieval_number}
    \end{minipage}
\end{figure}

We analyze how the number of retrieved passages (N) for each subquestion affects the QA performance. As illustrated in Figure \ref{figure_retrieval_number}, the sensitivity of N varies across various datasets. The results show that retrieving just a single passage in the in-domain dataset 2WikiMQA achieves optimal performance, which suggests that our retrieval method can effectively prioritizes relevant information above the irrelevant ones. 
Regarding to the out-of-domain datasets such as HotpotQA and IIRC, the performance improves with the increasing number of retrieved passages at first, and then tends to drop as more irrelevant information will be introduced by excessive retrieval. 
This indicates that retrieving additional appropriate information is usually beneficial to provide more clues for enhancing the QA performance in the out-of-domain scenarios.



\begin{figure}[t]
    \centering
    \begin{minipage}{0.5\textwidth}
        \includegraphics[width=\linewidth]{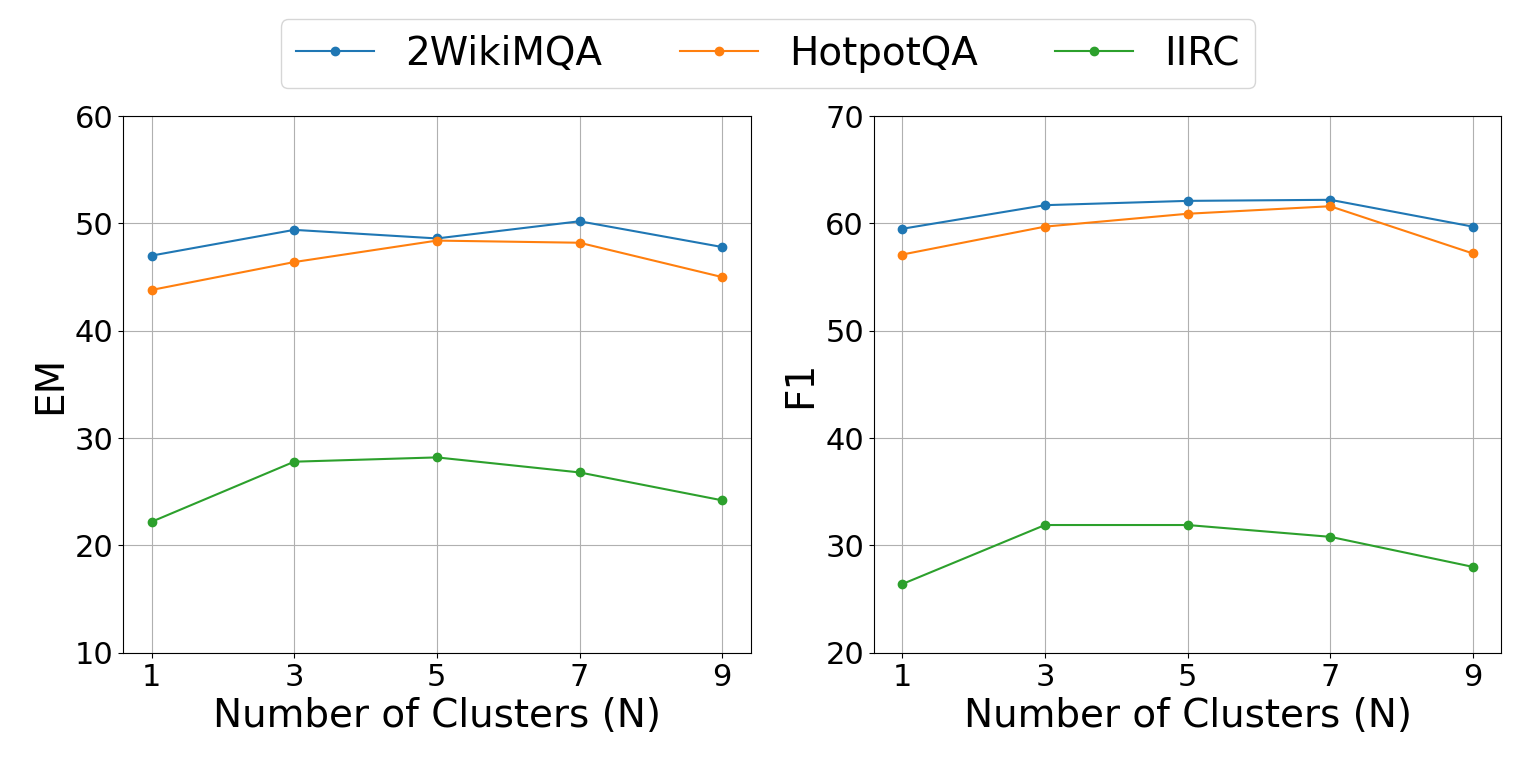}
        \caption{Impact of the number of clusters.}
        \label{figure_classes}
    \end{minipage}
\end{figure}

\paragraph{Impact of Clusters.}
The impact of the number of clusters is shown in Figure \ref{figure_classes}. We observe that the performance first increases with the growing number of clusters, and then declines. Specifically, too few clusters result in coarse-grained retrieval, which reduces the model’s ability to distinguish between different question spaces and relevant passages, and is prone to the semantic mismatching problem. Whereas, excessive clustering is also detrimental, as it will cause the semantically related questions to be unnecessarily separated into different spaces, which obstructs the learning of the common matching patterns and thereby diminishes the retrieval effectiveness. 
Overall, these findings indicate that an optimal cluster size balances the retrieval granularity and generalization. Too few clusters fail to capture fine-grained semantic distinctions, while too many clusters lead to semantic fragmentations, ultimately reducing the overall question answering performance.



\section{Conclusions}

In this paper, we propose a method to optimize the question semantic space for dynamic retrieval-augmented multi-hop question answering. By integrating three modules as \textit{QDM}, \textit{SDOM} and \textit{DPRM}, our method well bridges the semantic gaps between the question and helpful passages. Extensive experiments verify the effectiveness of each module of our method. In particular, our method outperforms the state-of-the-art baselines with significant improvements in both in-domain and out-of-domain settings, indicating the good generalization in unknown scenarios. Moreover, our method improves the retrieval accuracy, while maintaining high efficiency compared with the recent advanced retrieval methods. In the future, we will explore to extend our method to multilingual or multimodal settings, and investigate the effectiveness on more out-of-domain datasets.

\section*{Limitations}

Though \texttt{Q-DREAM} enhances the multi-hop question answering, its performance remains to be studied with other complex reasoning tasks. In addition, how to retrieve the long-tail knowledge for RAG remains to be studied.

\section*{Ethics Statements}
Language models may generate incorrect or biased information, especially when handling sensitive topics. While retrieval-augmented methods can help mitigate this issue, they do not fully eliminate the risk of biased or inappropriate content. Therefore, caution is necessary when deploying such systems in user-facing applications.

This work utilizes publicly available datasets (HotpotQA, 2WikiMQA, IIRC) that comply with academic licenses and do not contain personal or sensitive information. All models (e.g., ChatGPT, Llama2-7B) and training data were used in accordance with their respective terms of service. Our study does not involve human subjects or private data collection, thus avoiding risks related to consent or privacy. 

\section*{Acknowledgements}
This research is funded by the National Nature Science Foundation of China (No. 62477010 and No.62307028), the Natural Science Foundation of Shanghai (No. 23ZR1441800), Shanghai Science and Technology Innovation Action Plan (No. 24YF2710100 and No.23YF1426100) and Shanghai Special Project to Promote High-quality Industrial Development (No. 2024-GZL-RGZN-02008).





\bibliography{acl2025}
\end{document}